\documentclass[useAMS,usenatbib]{mn2e}
\usepackage{graphicx}
\usepackage{float}
\usepackage[usenames,dvipsnames]{xcolor}

\title[High-frequency iron K lags in Ark~564 and Mrk~335]{Discovery of high-frequency iron K lags in Ark~564 and Mrk~335}
\author[Kara et al.]{E. Kara$^{1}$\thanks{E-mail:
ekara@ast.cam.ac.uk}, A. C. Fabian$^{1}$, E. M. Cackett$^{2}$, P. Uttley$^{3}$, D. R. Wilkins$^{1}$, \newauthor{and A. Zoghbi$^{4,5}$}\\
$^{1}$Institute of Astronomy, Madingley Rd, Cambridge CB3 0HA\\
$^{2}$Department of Physics and Astronomy, Wayne State University, Detroit, MI 48201, USA\\
$^{3}$Astronmical Institute `Anton Pannekoek', University of Amsterdam, Postbus 94249, 1090 GE Amsterdam, the Netherlands\\
$^{4}$Department of Astronomy, University of Maryland, College Park, MD 20742, USA\\
$^{5}$Joint Space-Science Institute (JSI), College Park, MD 20742-2421, USA}

\begin{document}

\date{Accepted 2013 June 11.  Received 2013 June 11; in original form 2013 March 30}

\pagerange{\pageref{firstpage}--\pageref{lastpage}} \pubyear{2012}

\maketitle

\label{firstpage}

\begin{abstract}
We use archival {\em XMM-Newton} observations of Ark~564 and Mrk~335 to calculate the frequency dependent time-lags for these two well-studied sources.  We discover high-frequency Fe~K lags in both sources, indicating that the red wing of the line precedes the rest frame energy by roughly 100~s and 150~s for Ark~564 and Mrk~335, respectively.  Including these two new sources, Fe~K reverberation lags have been observed in seven Seyfert galaxies.  We examine the low-frequency lag-energy spectrum, which is smooth, and shows no feature of reverberation, as would be expected if the low-frequency lags were produced by distant reflection off circumnuclear material.  The clear differences in the low and high frequency lag-energy spectra indicate that the lags are produced by two distinct physical processes.  Finally, we find that the amplitude of the Fe~K lag scales with black hole mass for these seven sources, consistent with a relativistic reflection model where the lag is the light travel delay associated with reflection of continuum photons off the inner disc.\end{abstract}


\begin{table*}
\centering
\begin{tabular}{c|c|c|c|c|c|c|c}
\hline
Object & Obs. ID & Obs. Date & Duration (s) & Exposure (s) & Pile-up & Obs. Mode & Bkg. Reg. ('')\\
\hline
Ark~564 & 0670130201 & 2011-05-24 & 59500 & 59100 & yes & sw & 35\\
& 0670130301 & 2011-05-30 & 55900 & 55500 & no& sw& 35\\
& 0670130401 & 2011-06-05 & 63620 & 56820 & no& sw& 35\\
& 0670130501 & 2011-06-11 & 67300 & 66900 & yes& sw& 35\\
& 0670130601 & 2011-06-17 & 60900 & 60500 & no& sw& 35\\
& 0670130701 & 2011-06-25 & 64420 & 52620 & no& sw& 35\\
& 0670130801 & 2011-06-29 & 58200 & 57800 & yes& sw& 35\\
& 0670130901 & 2011-07-01 & 55900 & 55500 & yes& sw& 35\\
& 0206400101 & 2005-01-05 & 98956 & 98660 & no & sw & 35\\
Mrk~335 & 0306870101 & 2006-01-03 & 133000 & 120000 & no & sw& 35\\
& 0600540501 & 2009-06-11 & 80730 & 80700 & no & fw& 53\\
& 0600540601 & 2009-06-13 & 130300 & 109820 & no & fw& 35 \\
\hline
\end{tabular}
\caption{The {\em XMM-Newton} observations used in this analysis. Columns show the source name, the observation ID, the start date, duration of the observation, net exposure time after live time correction and background flares, whether pile-up corrections were made, the observation mode, and the size of the background region.}
\label{obs}
\end{table*}

\section{Introduction}
\label{intro}
The soft time lag(where soft band variations lag behind corresponding variations in the hard band) has revealed a new perspective through which to study the X-ray emission mechanisms of supermassive black holes. Since a tentative finding in Ark~564 \citep{mchardy07} and the first robust discovery in 1H0707-495 \citep{fabian09}, the soft lag has been detected in nearly twenty other Seyfert galaxies \citep[e.g. ][]{emmanoulopoulos11,demarco11,zoghbi11b,demarco13,fabian13,cackett13,zoghbi13}, and one black hole X-ray binary \citep{uttley11}.  Additionally, studying the high frequency soft lag as a function of energy \citep[as first done by ][]{zoghbi11}, has allowed us to directly compare the lag-energy spectrum with the time-integrated energy spectrum in order to understand which spectral component contributes to the lag at a particular energy \citep[see][for details]{kara13a}.  

Recently, the lag-energy spectrum has revealed the lag associated with the broad iron~K line.  This was first discovered in the bright Seyfert galaxy NGC~4151 by \citet{zoghbi12}.  In this work, they not only discovered the Fe~K lag, but also used a frequency-resolved approach to show that at lower frequencies (i.e. longer timescales), the rest frame energy of the Fe~K line lagged behind the continuum, while at higher frequencies (i.e. shorter timescales from a smaller emitting region), the longest lag comes from the red wing of the Fe~K line.  This is consistent with relativistic reflection off the inner accretion disc, where the reflected emission lags the continuum roughly by the light travel time between the corona and the accretion disc.

An alternative interpretation has been proposed by \citet{miller10}, in which the X-ray source is partially covered by optically thick clouds that are hundreds of gravitational radii from the source.  In this case, the long, low-frequency, hard lag is the reverberation between the source and the distant reflector, and the soft lag is simply the byproduct of taking the Fourier transform of the sharp-edged response function of clumpy reverberating material.  This alternative interpretation has not yet been shown to be able to explain the observed broad Fe~K lag at high frequencies.  Furthermore, the soft lag has also been observed in an X-ray binary, where there is no evidence for distant circumnuclear material \citep{uttley11}.   

Fe~K lags have so far been observed in a total of five sources \citep{zoghbi12,kara13a,kara13b,zoghbi13}. In this paper, we present the discovery of the Fe~K lag in two new sources, Ark~564 and Mrk~335.  The two sources are X-ray bright and highly variable, making them the ideal candidates for spectral timing studies.  The soft lag was first discovered in both of these sources by \citet{demarco13}.

Ark~564 ($z=0.0247$) was first observed with {\em XMM-Newton} in 2000/2001 at the beginning of the mission. In that short observation (10~ks of EPIC-pn exposure), the source was found to have a steep power law ($\Gamma \sim 2.5$) and large amplitude variability on short timescales \citep{vignali04}.  In 2005, the source was observed for 100~ks \citep{arevalo06b,papadakis07}. \citet{mchardy07} found that the power spectral density (PSD) of Ark~564 was well described by a power law with two clear breaks, similar to what is observed in Galactic black hole binaries in the soft state.  The PSD was also well fitted by a two Lorentzian model, suggesting that the variability originates from two discrete, localized regions. 

Most recently, Ark~564 was observed for 500~ks with {\em XMM-Newton}.  Using this long observation, \citet{legg12} confirmed a significant soft lag between the 0.3--1 keV and 4--7.5 keV bands.  This frequency-dependent time lag appeared narrower than in other sources, which led the authors to suggest a distant reflection origin.  However, we must note that the lag spectrum was not computed between usual the soft, 0.3--1~keV, and medium, 1--4~keV, energy bands as is done in other sources.

\begin{figure*}
\begin{center}
\includegraphics[width=\textwidth]{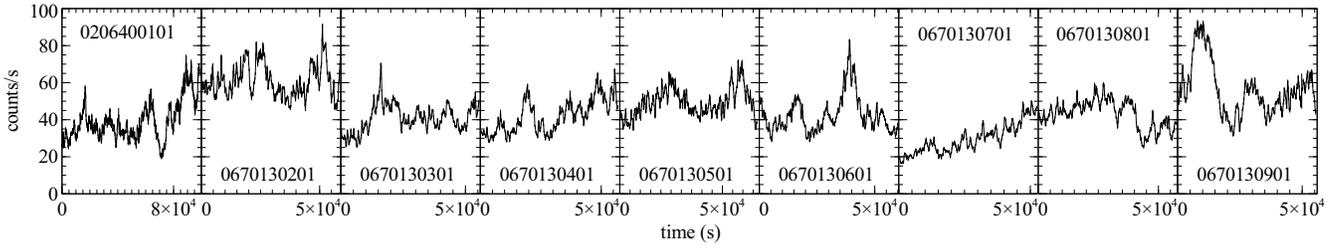}
\caption{Ark~564 broadband light curves from 0.3--10. keV in 200~s bins.  For the purposes of comparing the counts rate between different observations, the light curves were all generated with circular source extraction regions with 35 arcsec radii.  However, for the lag and spectral analysis, we accounted for pile-up by extracting annuli where needed. Note that the earlier observation (obsid 0206400101; leftmost panel) is longer, and is shown with a different x-axis scaling.}
\label{ark_lc}
\end{center}
\end{figure*}

Mrk~335 ($z=0.0258$) is a remarkably variable source that has been observed in several different spectral states.   Mrk~335 was observed for $\sim 130$~ks in 2006 when the source was in a high flux interval \citep{oneill07,larsson08}.  Using this long observation, \citet{arevalo08} studied the timing properties of Mrk~335 and found that the PSD is well described by a broken power law, with a break at $\sim 10^{-4}$~Hz, corresponding to the frequency at which there was a sharp cut-off in the hard lag.  They concluded that the frequency dependent time lags were consistent with fluctuations propagating through the accretion flow.  Recently, \citet{gallo13}, using a 200~ks {\em XMM-Newton} obtained in 2009 \citep{grupe12}, showed that spectral and timing properties of this 200~ks observation were consistent with blurred reflection from an accretion disc around a rapidly spinning black hole ($a>0.7$).

The paper is organized as followed: The observations and data reduction used in this analysis are described in Section~\ref{obs_sec}. We review the Fourier method for calculating the lag in Section~\ref{fourier}, and present the results in Section~\ref{results}.  Finally, we discuss the results in Section~\ref{discuss}, and show how these results fit in with the growing sample of AGN with Fe~K lags.

\begin{figure}
\begin{center}
\includegraphics[width=6cm]{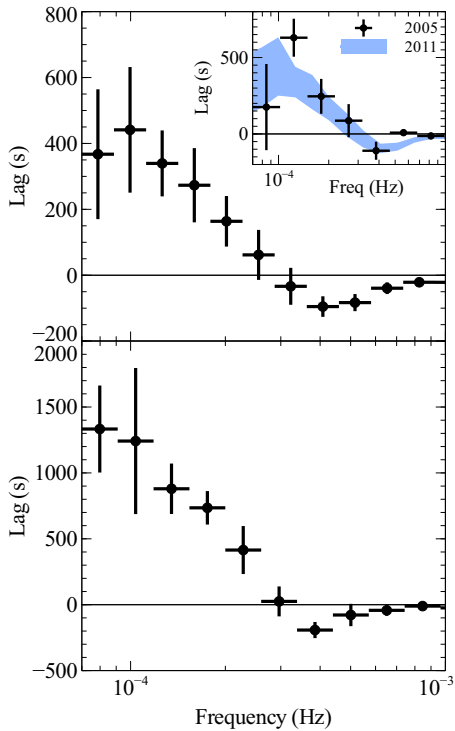}
\caption{The lag as a function of temporal frequency for Ark~564. The lag is measured between 0.3--1~keV and 1.2 -- 4~keV ({\em top}) and between 0.3 -- 1~keV and 4 -- 7.5~keV ({\em bottom}).    
The amplitudes of the soft and hard lags are much larger in the bottom panel. However, we note that the frequency of the soft lag is the same for both energy ranges.  The inset shows the 0.3--1~keV to 1.2--4~keV lag using only the 100~ks observation from 2005 (black points) overlaid with the 1-$\sigma$ contours from the 2011 lag-frequency spectrum in blue. The 2005 soft lag is not significant, but is consistent with the new observations.}
\label{ark_lagfreq}
\end{center}
\end{figure}

\section{Observations and Data Reduction}
\label{obs_sec}

For the analysis of these two sources, we use all of the archival data from the {\em XMM-Newton} observatory \citep{jansen01}, shown in Table~\ref{obs}. The data for both sources were reduced in the same way using the {\em XMM-Newton} Science Analysis System (SAS v.11.0.0) and the newest calibration files.  We focus on the data from the EPIC-pn camera \citep{struder01} because of its faster readout time and larger effective area at high energies. The MOS data yield consistent results, but as the addition of the MOS data does not change or improve the overall lag spectra, they were not included in the analysis.  

The data were cleaned for high background flares, and were selected using the conditions {\sc pattern} $\le 4$ and {\sc flag}~$==0$.  The source spectra were extracted from circular regions of radius 35 arcsec centered on the maximum source emission. If pile-up was present, then an annular region was used to exclude the innermost source emission.  Pile-up was an issue for some observations in Ark~564, and for consistency, we chose the same size excision radii used in \citet{legg12}.  The majority of the observations were taken in prime small window imaging mode, and for this mode, the background spectra were chosen from circular background regions, also of 35 arcsec radius. For observations taken in full window imaging mode (only the 2009 observations of Mrk~335), the background regions were made as large as possible, sometimes as large as twice the radius of the source region.  The background subtracted light curves were produced with the tool {\sc epiclccorr}. The light curves were all binned with 10~s bins.

\section{The Fourier Method}
\label{fourier}

To compute the time lags, we use the Fourier technique outlined in \citet{nowak99}.  This gives us time lags as a function of temporal frequency (i.e., timescale$^{-1}$).

We start by generating light curves in different energy bands consisting of $N$ observations in 10~s bins (i.e., dt=10~s).  The frequency range is limited at low frequencies by the length of the observation.  The high frequency cutoff is strictly the Nyquist frequency at $f=1/(2 dt)$, however we are dominated by Poisson noise at frequencies well below the Nyquist frequency.

We take the discrete Fourier transform of each light curve.  In this example, we will find the time delay between a soft band light curve, $s(t)$ and a hard band light curve, $h(t)$. The discrete Fourier transform of the soft band light curve is
$$\tilde{S}(f) = \frac{1}{N}\sum_{j=0}^{N-1} s(t_j) e^{-2 \pi ift_j/N},$$
where $N$ is the number of time bins in the light curve, $s(t)$, and the frequency, $f=j/(N dt)$.
The soft band Fourier transform can be written in the phasor form as the product of its amplitude and complex exponential phase
$$\tilde{S}(f)=| \tilde{S}(f) | e^{i\theta_{s}}.$$
We take the Fourier transform of the hard band light curve, and then take its complex conjugate
$$\tilde{H}^{\ast}(f)=| \tilde{H}(f) | e^{-i\theta_{h}},$$
which reverses the sign of the phase.
The product of $\tilde{S}(f)$ with the complex conjugate of the hard band , $H^{\ast}(f)$, is known as the cross product $\tilde{C}(f)$ and is written, 
$$\tilde{C}(f) = \tilde{H}^{\ast} \tilde{S} = | \tilde{H} | | \tilde{S} | e^{i(\theta_s-\theta_h)}.$$ 
This gives the phase difference between the soft and the hard band.  The overall Fourier phase lag, $\phi(f)$ is the phase of the average cross power spectrum. That is, 
$$\phi(f) = \mathrm{arg}[\langle \tilde{C}(f) \rangle].$$
The phase lag is then converted back into time to give us a frequency dependent time lag between the two light curves:
$$\tau(f) \equiv \frac{\phi(f)}{2\pi f}.$$

It is important to note that the Fourier phase is defined on the interval $(-\pi,\pi)$, and therefore we are subject to phase wrapping, which causes the amplitude of the lag to change sign \citep{nowak99}. In other words, a constant lag of amplitude $\tau$ will change sign when $\phi=-\pi,\pi$ at the frequency $f=1/(2\tau)$ \citep[See][for a more detailed discussion of phase wrapping]{wilkins13}.

\begin{figure}
\begin{center}
\includegraphics[width=7cm]{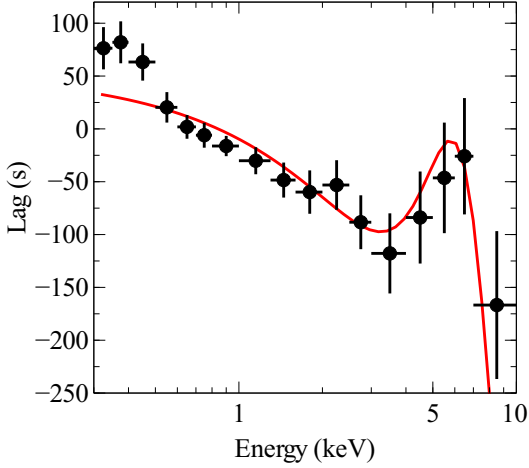}
\caption{The lag-energy spectrum of Ark~564 in the frequency range $[3.2-5.2] \times 10^{-4}$~Hz.  The lag-energy profile shows a peak at the energy of the Fe~K line, and is similar to the soft lag-energy profiles seen in 1H0707-495 and IRAS~13224-3809. The red line shows the best fit linear model with an additional Gaussian at 6.4~keV.}
\label{ark_lagen}
\end{center}
\end{figure}

\begin{figure}
\begin{center}
\includegraphics[width=7cm]{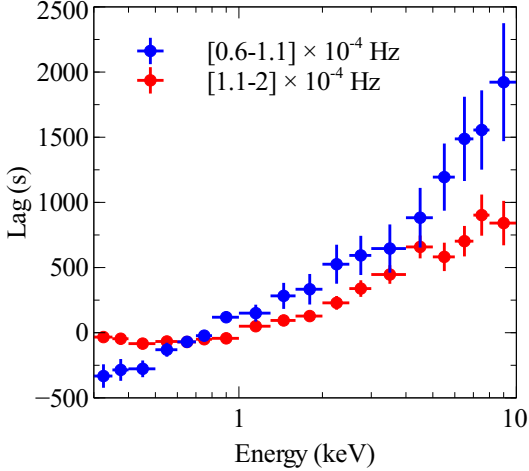}
\caption{The low-frequency lag-energy spectrum of Ark~564 in the low frequency range $[0.6-1.1] \times 10^{-4}$~Hz (blue) and $[1.1-2] \times 10^{-4}$~Hz (red).  The profile is smooth and very different from the high-frequency lag-energy spectra.  }
\label{ark_hardlagen}
\end{center}
\end{figure}

\begin{figure}
\begin{center}
\includegraphics[width=7cm]{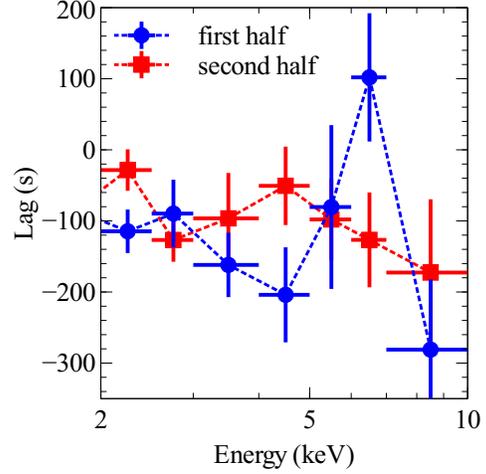}
\caption{The lag-energy spectra of Ark~564 for the first half of the observation (obsid 201-501; {\em blue}) and the second half (obsid 601-901; {\em red}). The first half lag-energy spectrum is shown at $\nu=[3.4-4.5] \times 10^{-4}$~Hz, and the second half is shown at a slightly larger frequency range, $\nu=[3.2-5.2] \times 10^{-4}$~Hz.  The first half at lower frequencies clearly shows a narrower peak at 6-7~keV, while the second half at higher frequencies shows a broader feature peaking at 4-5~keV.}
\label{ark_redblue}
\end{center}
\end{figure}

\subsection{The transfer function}
\label{transfunc}

An important step in understanding the origin of the lag, is to characterize the linear transformation that relates two delayed light curves. This function, known commonly in signal processing as the impulse response function $t_{r}(\tau)$, relates the continuous hard and soft band light curves such that
$$s(t) = \int\limits_{-\infty}^{\infty}t_{r}(t-\tau)h(\tau)d\tau.$$
The Fourier transform of the impulse response function is called the transfer function, $\tilde{T_r}(f)$ \footnote{The term `transfer function' is often used in astronomy to describe the linear transformation in the time domain.  However, according to conventional signal processing terminology, the `transfer function' only refers to this linear transformation in the {\em frequency} domain. Therefore, strictly speaking, the `transfer function' as typically known in astronomy, is actually called the `impulse response function'. In this paper, we will adopt this correct terminology.}. The linear transformation between the soft and hard bands in the frequency domain is
$$\tilde{S}(f)=\tilde{T_r}(f)\tilde{H}(f).$$
Given this relation between the soft and hard bands, the cross spectrum can just be written
$$\tilde{C}(f)=|\tilde{H}(f)|^2  \tilde{T_r}(f)$$
Therefore, the lag is derived from the phase of the transfer function.  Theoretically, we should be able to directly measure the transfer function over a given frequency range by dividing the cross spectrum by the hard band power spectrum.  In practice, however, statistics are generally too low to allow a direct and unique measure of the transfer function using this method because calculating the transfer function requires a deconvolution of the signal in the two energy bands, which is known to enhance the noise level \citep[See Section 13.1 of][for further discussion of the deconvolution]{press92}.  Such analysis may be possible with future missions, such as {\em LOFT} or {\em Athena+}.

\label{mrk}
\begin{figure*}
\begin{center}
\includegraphics[width=\textwidth]{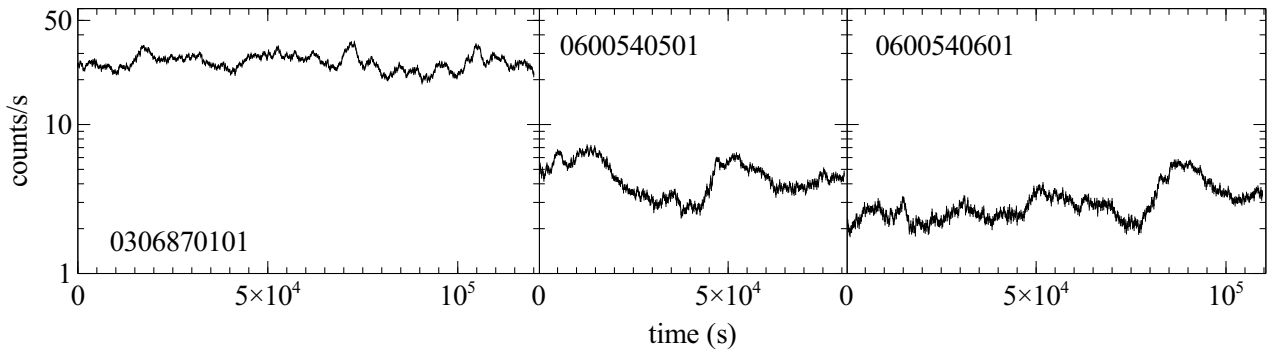}
\caption{Mrk~335 broadband light curve from 0.3--10.~keV in 200~s bins. Note the logarithmic y-axis.  The 2006 observation ({\em left}) is $\sim 10$ times brighter than the 2009 observations ({\em middle, right}).}
\label{mrk_lc}
\end{center}
\end{figure*}

A common approach to studying the response function is to model the impulse response function \citep[e.g.][]{reynolds99}, and compare the phase lag of the response function to the observed lags.  This technique has been done by \citet{wilkins13} using general relativistic ray tracing simulations from the source, to the accretion disc and finally to the observer.   In this inner disc reflection model, the hard lag is produced by a different mechanism (i.e. propagation effects through corona that cause the soft photons to respond before the hard photons).  Therefore, the hard lags require a different response function.  \citet{miller10} also used this technique of assuming a response function that fits the observed lag. They use a top hat response function of length $\Delta \tau$ where $\Delta \tau$ corresponds to the maximum light travel delay between the X-ray source and a line-of-sight distant reflector.  The sharp edge in this response function was shown to reproduce the high-frequency soft lag observed in a 500~ks observation of 1H0707-495.

\subsection{Computing the lag-energy spectrum}

The lag-energy spectrum is an important tool in directly comparing the lag with the time-integrated energy spectrum. We use the lag-energy spectrum as a way to see what parts of the spectrum contribute to the lag at specific frequencies.

 The lag-energy spectrum is computed by measuring the frequency-dependent lag (described above) between the light curves of narrow energy bins ($\Delta E/E \sim 0.12 $) and a reference band light curve \citep[e.g.][]{zoghbi11}.  The choice of reference band does not change the shape of the lag-energy spectrum, but it does change where the zero-point occurs.  For this analysis, we chose the reference band to be the entire energy range, from 0.3--10. keV, excluding the small energy bin itself, so as not to have correlated errors \citep[as discussed further in ][]{zoghbi13}.

We note, therefore, that the interesting lag is not the absolute lag amplitude, but rather the relative lag between the energy bins.  We read the lag-energy spectrum from bottom to top, the lower-valued/more negative lag precedes the higher-valued lag.

\section{Results}
\label{results}

\subsection{Ark~564}
\label{ark}

\begin{figure}
\begin{center}
\includegraphics[width=7cm]{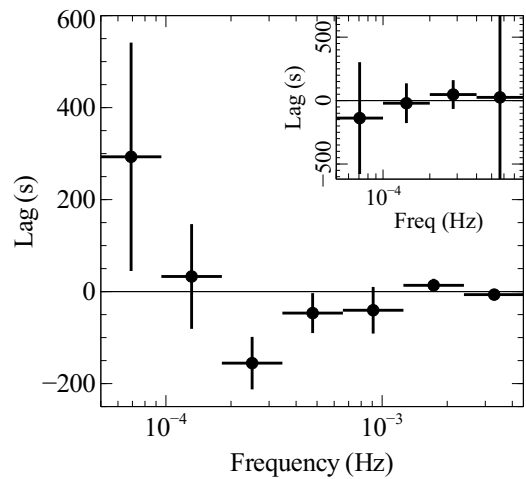}
\caption{The Mrk~335 lag as a function of temporal frequency between 0.3--0.8~keV and 1--4~keV for the 2006 observations.  The inset shows the lag as a function of frequency for just the 2009 low flux observations.  No lag was detected, and therefore we did not use these observations in the lag-energy analysis.}
\label{mrk_lagfreq}
\end{center}
\end{figure}

The light curves of Ark~564 in Fig.~\ref{ark_lc} show a bright and highly variable source over the 500~ks observation, making it a good candidate for searching for time lags. Fig.~\ref{ark_lagfreq} shows the lag as a function of temporal frequency, using the technique described in Section~\ref{fourier}. The top panel shows the lag between the soft band (0.3 -- 1 keV) and a middle band (1.2--4 keV).  The bottom panel shows the lag between the soft band and a hard band (4--7.5~keV), as first shown in \citet{legg12}. Note the different scales on the y-axis.  The two panels show different lag amplitudes, but the frequency of the soft lag is the same, independent of the amplitude of the hard lag.  We note that the transition from positive to negative lag is not steep, as would be expected from an impulse response function with a sharp edge.

The amplitude of the lag between the soft and medium band is $-95 \pm 31$~s at $[3.5-4.5] \times 10^{-4}$~Hz.  Assuming a black hole mass of $1.7 \times 10^6 M_{\odot}$ \citep{denney09}, the amplitude and frequency of the soft lag are consistent with the black hole mass scaling relation of \citet{demarco13}.

The inset in the top panel of Fig.~\ref{ark_lagfreq} shows the lag between 0.3--1~keV and 1.2--4~keV using the 100~ks observation from 2005 in black points, overlaid with the 1-$\sigma$ contours of the 2011 lag-frequency spectrum in blue.  The 2005 observation was taken when the source was in a similar flux regime.  While the soft lag is not significant, it is consistent with the newer observations.  The amplitude of the lag at $[3.5-4.5] \times 10^{-4}$~Hz is $-140 \pm 120$~s, which is within error of the 2011 lag at the same frequency. The lag analysis for these data was first presented in \citet{mchardy07}, where they fit a simple function across the entire high-frequency range.  Using this technique, they estimated the amplitude of the soft lag between 0.5--2~keV and 2--8.8~keV to be $-11.0 \pm 4.3$~s.  The small amplitude and error of this lag is likely due to fitting for a constant lag across all the high-frequencies, even where we see the lag is zero.  Including this 100~ks observation into our analysis does not change the average lag-frequency spectrum or the lag-energy spectrum, and so we choose to present the results from only the four orbits from 2011, in order to make a direct comparison with the results presented in \citet{legg12}.

\begin{figure}
\begin{center}
\includegraphics[width=7cm]{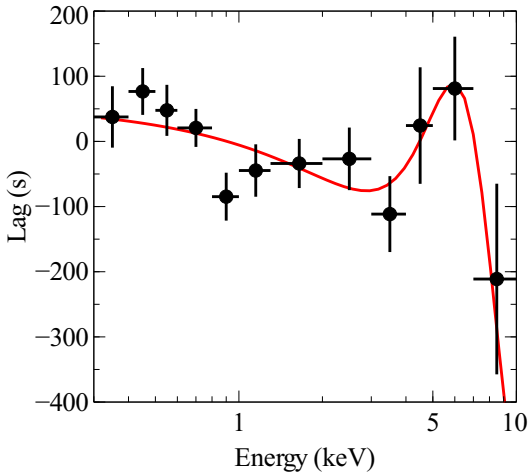}
\caption{The Mrk~335 lag-energy spectrum in the frequency range $[1.9-6.7] \times 10^{-4}$~Hz. Only the data from 2006 were used as no lag was present in the 2009 observation.  The red line shows the best fitting linear model plus a Gaussian at 6.4~keV.} 
\label{mrk_lagen}
\end{center}
\end{figure}

The high-frequency lag-energy spectrum ($[3.2-5.2] \times 10^{-4}$~Hz) in Fig.~\ref{ark_lagen} has the same general profile as the high-frequency lag-energy spectra seen in 1H0707-495 and IRAS~13224-3809.  The high energy lag spectrum is very similar to these other two sources, with a clear local maximum at 6--7~keV, the energy of the Fe~K line, and a dip at 3--4 keV.  At 0.3--0.5~keV, we see a larger delay than in the other two sources, which may be indicative of ionization differences or more contribution from reprocessed black body emission from the irradiated accretion disc that causes a larger contribution of delayed emission relative to the direct emission.

As a check, we show the statistical significance of a model Gaussian to the feature at 6.4~keV.  We fit a function to the data, $\tau(E) = a + bE + c e^{-\left(\frac{E-6.4}{d}\right)^2}$, shown as the red line in Fig.~\ref{ark_lagen}.  This function was modelled on the lag-energy spectra of 1H0707-495 and IRAS~13224-3809, where there are clear indications of a Fe~K feature. We compared this model to one of just a line without the additional Gaussian. Comparing these two nested models yields an $F$-statistic, i.e. $\Delta\chi^2/\chi^2_{\nu}$, of 5.5.\footnote{The F-test approach of testing for additive nested components where only the normalization is changed is standard for X-ray spectral fitting. Since we ensure that the normalization can be negative, we avoid the problem of fitting close to the parameter space boundary \citep{protassov02}}  Therefore, the Gaussian model is preferred with 98.2\% confidence.  We notice, however, that in this source, the 0.3--0.5~keV band diverges from a linear relation.  If we avoid the soft excess, and only fit the functions between 0.5--10~keV, the Gaussian model is preferred with 99.94\% confidence.


\begin{figure}
\begin{center}
\includegraphics[width=7cm]{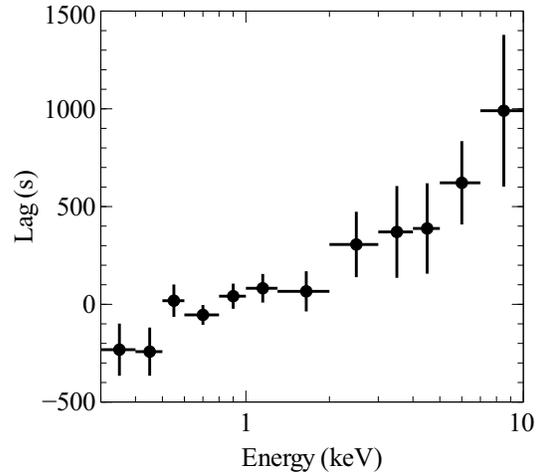}
\caption{The low-frequency lag-energy spectrum of Mrk~335 in the low frequency range $[0.4-1.6] \times 10^{-4}$~Hz.  Like Fig.~\ref{ark_hardlagen} the profile appears smooth and very different from the high-frequency lag-energy spectra.  }
\label{mrk_hardlagen}
\end{center}
\end{figure}

In Fig.~\ref{ark_hardlagen}, we show the low-frequency hard lag dependence on energy for two slightly different frequency ranges.  The lowest frequency range ($[0.6-1] \times 10^{-4}$~Hz in blue) shows a simple power-law shape, while the slightly higher frequency range ($[1.1-2]\times10^{-4}$~Hz in red) is flat until 1~keV, when it turns upwards and follows the same power-law behavior.

\begin{figure*}
\begin{center}
\includegraphics[width=\textwidth]{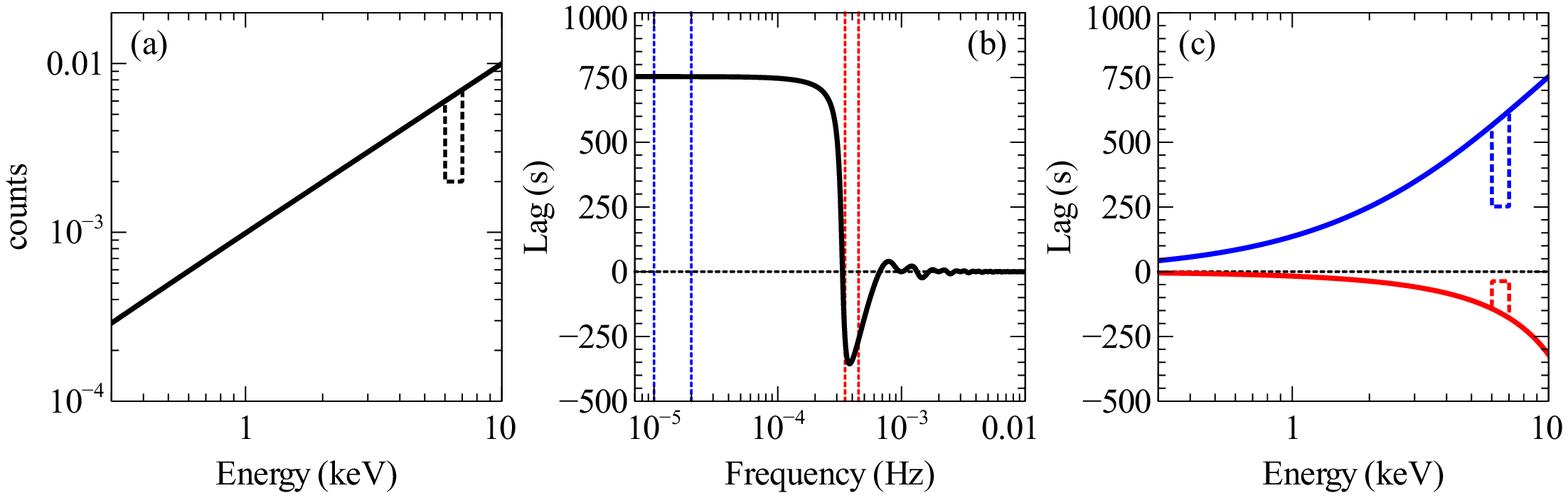}
\caption{Model of the lag-energy spectra given a simple top-hat response function for two test cases: (1) where the reflection fraction increases steadily with energy (solid line), and (2) where the reflection fraction increases steadily with energy except for a demonstrative dip at 6--7~keV (dotted line).  We show the reflection spectrum (a), the lag-frequency spectrum (b) and the low and high frequency lag-energy spectrum (c) shown in blue and red, respectively.}
\label{sim}
\end{center}
\end{figure*}

In the first reported results on this 500~ks observation of Ark~564, \citet{legg12} showed possible weak non-stationarity in the light curves from the first half of the observation (obsid 201-501) and the second half (obsid 601-901), taken a few weeks later.  They reported a different frequency dependence of the soft lag and the PSD. Here we report a slight difference in the lag-energy spectrum between the first and second halves of the observation (Fig.~\ref{ark_redblue}).  The lag-energy spectrum of the first half of the observation ($f=[3.4-4.5] \times 10^{-4}$~Hz) shows a narrow lag feature peaking at 6--7~keV.  The lag-energy spectrum of the second half of the observation covers a slightly larger range up to higher frequencies, $f=[3.2-5.2] \times 10^{-4}$~Hz, and shows a more constant behavior, possibly with a peak at 4--5~keV.  We note that the chosen energy bin size can affect the amplitude of the lag (i.e. averaging over larger energy bins will inevitably decrease the structure peaking at 6--7~keV).  Nonetheless, using this finer binning, we find that the first half peaks at higher energies than the second half. The energy dependence seen here is reminiscent of the frequency-dependent lags in NGC~4151 that show a more narrow peak at low frequencies, and a broader feature at high frequencies \citep{zoghbi12}.  No difference was found in the hard lag-energy spectra between the first and the second halves of the observation.

\subsection{Mrk~335}
Mrk~335 was observed in 2006 when it was very bright. Three years later, the source was observed again, this time while in a lower flux interval, with an average flux around 10 times less.  The light curves in Fig.~\ref{mrk_lc} show the 300~ks of observations taken in these dramatically different flux intervals.

We compute the lag as a function of frequency between 0.3--0.8~keV and 1.--4.~keV for the 2006 observation of Mrk~335 (Fig.~\ref{mrk_lagfreq}).  The most negative lag is $-155 \pm 57$~s at $f \sim 2 \times 10^{-4}$~Hz. We attempted to measure a lag in the lower flux observations from 2009, but no lag was found (inset of Fig.~\ref{mrk_lagfreq}).  The combined lag from all three observations \citep[as shown in a previous lag measurements of Mrk~335 in][]{demarco13} is not significantly different than the lag presented in Fig.~\ref{mrk_lagfreq} because (1) the flux of the 2006 observation is much higher and therefore dominates the combined spectrum and (2) there is a constant zero lag measured in the low-flux observations, and therefore the shape of the lag-frequency spectrum does not change.  In the low-flux observations from 2009, the lag in the same frequency range $[1.9-7.6]\times 10^{-4}$~Hz is measured to be $-17.6 \pm 81.0$~s.

We isolate the frequencies $[1.9-6.7] \times 10^{-4}$~Hz in the high flux observation from 2006 in order to look at the lag dependence on energy (Fig.~\ref{mrk_lagen}). We find a lag of amplitude $\sim$~150~s at the energy of the Fe~K line. Using the same approach as for Ark~564, we fit the high-frequency lag-energy spectrum with two simple models: a line and a line plus a Gaussian at 6.4~keV (shown as the red line in Fig.~\ref{mrk_lagen}).  The line plus Gaussian model is preferred with $\sim$~92\% (from an $F$-statistic of 3.6). Interestingly, even though Mrk 335 is known to have a stronger iron line than Ark~564, the detection of the Fe~K lag is weaker in Mrk~335.

For completeness, we show the low-frequency lag-energy spectrum ($[0.4-1.6] \times 10^{-4}$~Hz) for the 2006 observation of Mrk~335 (Fig.~\ref{mrk_hardlagen}).  Similar to the low-frequency lag-energy spectrum of Ark~564 in  Fig.~\ref{ark_hardlagen}, we find the low-frequency lag increases with energy. Though the error bars are larger for this source, it also appears to have a featureless low-frequency lag-energy spectrum.

\section{Discussion}
\label{discuss}

The X-ray time lags from Seyfert galaxies have been explained mainly by two physical interpretations: relativistic reflection off the inner accretion disc, and distant reflection from circumnuclear clouds.  Both interpretations propose reflection of continuum photons to explain the broad iron line and the X-ray time lags.  In the relativistic reflection model, the {\em short, high-frequency lags} are interpreted as reverberation between the continuum-emitting corona and the reflecting accretion disc \citep{fabian09}, and the low-frequency lags are produced by a different mechanism, consistent with propagating fluctuations in the disc that are transferred to the corona \citep{kotov01,arevalo06}.  In the partial covering model, it is the {\em long, low-frequency lags} that are interpreted as reverberation, now between the X-ray source and the distant reflector.  In this model, the high-frequency lag is a mathematical artefact due to phase wrapping \citep{miller10}.  

The high-frequency Fe~K lags, as detected here in Ark~564 and Mrk~335 (Fig.~\ref{ark_lagen} and Fig.~\ref{mrk_lagen}, respectively), are key to breaking the degeneracies between these two models.  The broad Fe~K line is an indicator of reflection in both models, however, we only observe a Fe~K reflection feature in the high-frequency lags, {\em not in the low-frequency lags} (i.e. Fig.~\ref{ark_hardlagen} and Fig.~\ref{mrk_hardlagen}).  In the case of partial covering, where the low-frequency lags are associated with reflection, it is not clear why the Fe~K reflection feature is only seen at high frequencies. Furthermore, it is not clear how to reproduce the Fe~K structure at high-frequencies, if it is simply due to phase wrapping.

Variations of the partial covering model have been proposed to explain the lag-frequency spectrum of Ark~564 \citep[using a small system of absorbing clouds at $100 r_g$ or less, modelled as a single top hat response function;][]{legg12} and for 1H0707-495 \citep[using an extended wind from tens to hundreds of gravitational radii, modelled as several top hat functions of different widths][]{miller10}.  While these different partial covering models can explain the lag-frequency spectra, neither has been shown to self-consistently explain the low and high frequency lag-energy spectra.

\subsection{Partial covering view of lag-energy spectra}

Here we calculate the expected low and high frequency lag-energy spectra given the simple single top-hat response function, similar to the one concluded in \citet{legg12}. 
In this work, the authors conclude that the lag-frequency spectrum between the soft band (0.3--1~keV) and the hard band (4--7.5~keV) can be explained as a single top-hat response function in the hard band lagging behind a delta function in both bands.  The top panel of Fig.~\ref{ark_lagfreq} shows the lag-frequency spectrum between the soft (0.3--1~keV) and medium bands (1--4~keV), which has a shorter low-frequency lag than the low-to-high lag-frequency spectrum, but has the same turnover frequency from hard to soft lags. This has been explained in the partial covering scenario of distant absorbers as a dilution effect \citep{miller10}.  In this case, the source photons scatter through a medium whose opacity decreases with increasing energy, and therefore the fraction of scattered photons increases with energy (as demonstrated by the top-hat response function whose amplitude increases with energy). This causes the amplitude of the low-frequency hard lag to increase with energy, but does not change the turnover frequency.

We illustrate this partial covering scenario in Fig.~\ref{sim} by computing the lag-frequency spectrum and the low and high frequency lag-energy spectra for a single top-hat response function whose amplitude increases with energy. This assumes that the reflector's response to the direct emission is uniform over a distribution of time delays, with a width of $\Delta \tau$, and centered at $t_0$.  For this demonstration, we choose $\Delta \tau=t_0=100$~s.  The solid line in the Panel~(a) shows the energy spectrum of the scattered light, steadily increasing with energy. We also include the case where the amplitude of the response function is not simply steadily increasing with energy (as shown by the dotted line with a clear feature at 6--7~keV). Panel~(b) depicts the frequency-dependent lag between the direct emission and the reflected emission at 10~keV.  Here, the soft lag is an artefact of taking the argument of the Fourier transform of the sharp-edge response function.  Solid blue and red lines in Panel (c) show the resulting lag-energy spectrum for low ($[1-2]\times10^{-5}$~Hz) and high frequencies ($[3.5-4.5]\times 10^{-4}$~Hz), respectively.  The increase in low-frequency lag with energy is a direct result of an increasing amount of dilution (i.e. an increasing fraction of scattered light).  The high-frequency lag, as a relic of the hard lag, is therefore also affected by dilution, and we see a mirror image in the high-frequency lag-energy spectrum.   We note that if the lag-energy dependence is due to a change in the width of the response function (rather than just a change in the amplitude as in the case presented here), the frequency turnover from hard to soft lags changes, and therefore causes a different shape to the high-frequency lag-energy spectrum that is not a mirror image of the hard lags.  Given that in Ark~564 the turnover frequency from hard to soft lags does not change, we can rule out the possibility of a change in the width of the response function, but even still, this scenario does not reproduce the observed peak at 6.4~keV.

We have shown here that the response function proposed in \citet{legg12} can explain the low-frequency lag-energy spectrum (Fig.~\ref{ark_hardlagen}), but cannot reproduce the Fe~K feature seen clearly in the high-frequency lag-energy spectrum (Fig.~\ref{ark_lagen}).  {\em We conclude that the high-frequency lag must have a different response function (and a different physical mechanism) from the low-frequency lag.}

\begin{figure}
\begin{center}
\includegraphics[width=7cm]{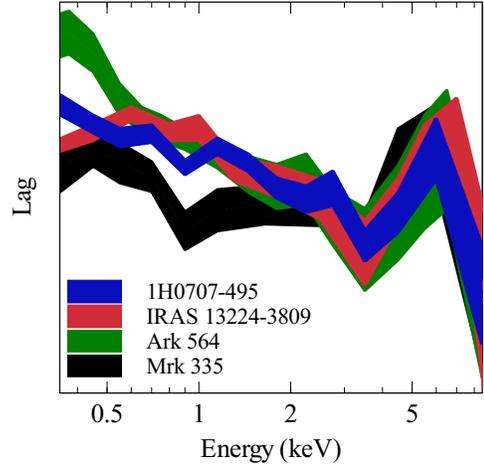}
\caption{The high frequency lag-energy profiles for 4 of the 7 sources that exhibit Fe~K lags, including 1H0707-495 ({\em blue}), IRAS~13224-3809 in at low fluxes ({\em red}), Ark~564 ({\em green}) and Mrk~335 ({\em black}).  As the zero-point is arbitrary, the lags have been scaled to the 3--4~keV band.  The Fe~K lag profile is the same in all four sources, but the lags at the soft excess do not share the a common shape.} 
\label{lagen_stack}
\end{center}
\end{figure}

\subsection{The growing sample of Fe~K lags}

High-frequency Fe~K lags have now been discovered in seven Seyfert galaxies.  Fig.~\ref{lagen_stack} shows the lag-energy spectrum for four of the seven sources (1H0707-495, IRAS~13224-3809, Ark564 and Mrk~335) with 68\% confidence contours.  Since the zero-point is arbitrary and these sources have different lag amplitudes, the lag-energy spectra have been scaled to the 3--4 keV band.  We focus on the overall shape, rather than the value of the lag.  The four sources show very similar Fe~K features, with a dip at 3--4~keV, a peak at 6--7~keV, and the most negative point at 8--10~keV.  At lower energies, however, the lag-energy spectra diverge.  There could be many reasons for this, including differences in iron abundance, ionization parameter, spectral index, or black body temperature.  Lags reveal a completely independent way of looking at the soft excess, and, with further work, could prove helpful in understanding the complexities of this energy band.

\begin{figure}
\begin{center}
\includegraphics[width=7cm]{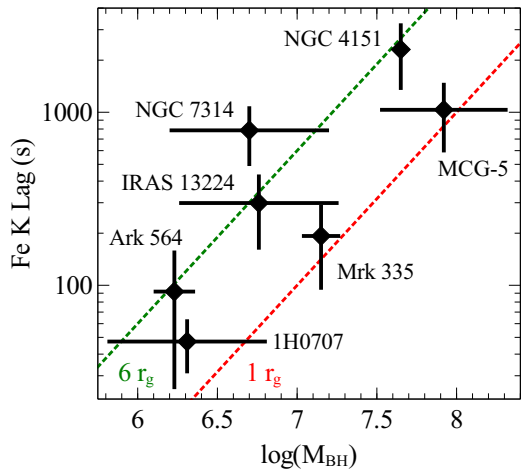}
\caption{The amplitude of the Fe~K lag vs. $M_{\mathrm{BH}}$ for the seven Seyferts with detected Fe~K reverberation lags.  See Table~\ref{sources} for details about the sources. The light crossing time at 1~$r_{\mathrm{g}}$ and 6~$r_{\mathrm{g}}$ as a function of mass are shown as red and green lines, respectively.}
\label{lag_mass}
\end{center}
\end{figure}

We plot the Fe~K lag amplitude vs. black hole mass for the seven source (Fig.~\ref{lag_mass}).  The Fe~K lag is measured between the 3--4~keV dip and the peak at 6--7~keV.  The black hole mass measurements were taken from the literature. Details can be found in Table~\ref{sources}.  There is roughly a linear relationship between the amplitude of the lag, and black hole mass, as expected by relativistic reflection. The light crossing time at 1~$r_{\mathrm{g}}$ and 6~$r_{\mathrm{g}}$ as a function of mass are shown as red and green lines, respectively, indicating that we are probing very close to the central black hole.  We note that we do not account for the effect of dilution, and therefore, the intrinsic lag amplitudes are likely to be slightly larger than the observed lag. Also, we do not include the effect of the Shapiro delay, which is likely important close to the black hole \citep{wilkins13}.  By comparing the lag at these high energies, we avoid complexities caused by the soft excess, but what is interesting, is that we still see the same mass scaling relation as found for the lags of the soft excess \citep{demarco13}. This points to a common origin of both the soft excess and the broad Fe~K lag at small distances from the X-ray corona.

\begin{table}
\centering
\begin{tabular}{c|c|c}
\hline
Object & log($M_{\mathrm{BH}}$) & $M_{\mathrm{BH}}$ reference \\
\hline
Ark~564 & $6.23 \pm 0.13$ & \citet{denney09}\\ 
1H0707-495 & $6.31 \pm 0.5$ & \citet{bian03}\\
NGC~7314 & $6.70 \pm 0.5$ & \citet{schulz94}\\
IRAS~13224-3809 & $6.76 \pm 0.5$ & \citet{zhou05}\\
Mrk~335 & $7.15 \pm 0.12$ & \citet{peterson04}\\
NGC~4151 & $7.65 \pm 0.03$ & \citet{bentz06}\\
MCG-5-23-16 & $7.92 \pm 0.4$ & \citet{oliva95} \\
\hline
\end{tabular}
\caption{Details of the black hole mass for the seven sources with an observed Fe~K lag.}
\label{sources}
\end{table}

Lastly, we comment briefly on the apparent weak non-stationarity in the light curves \citep[as noted by][through the lag-frequency spectrum and PSD]{legg12}.  Fig.~\ref{ark_redblue} shows that the first half of the 500~ks observation peaks at the rest frame of the Fe~K line, presumably from a larger emitting region, while at slightly higher frequencies, the second half shows a broader lag peaking at 4--5~keV, the red wing of the Fe~K line from a smaller emitting region.  While the non-stationarity of the light curves is not currently well understood, it does provide tantalizing evidence for further behavior that will be accessible either through much deeper observations or future observatories.

The case for relativistic reflection continues to grow with recent discoveries of the Fe~K reverberation lag \citep{zoghbi12} that is now seen in seven sources, flux-dependent reverberation lags \citep{kara13b}, the black hole mass scaling relation with lag \citep{demarco13}, and reverberation lags in a black hole X-ray binary \citep{uttley11} and possibly a neutron star \citep{barret13}.  The work presented here on Ark~564 and Mrk~335 support a relativistic reflection model and are not consistent with partial covering.  The study of X-ray reverberation lags is quickly developing, and is revealing a new perspective through which to probe strong gravity.   

\section*{Acknowledgements}

This work is based on observations obtained with {\em XMM-Newton}, an ESA science mission with instruments and contributions directly funded by ESA Member States and NASA.  EK thanks the Gates Cambridge Scholarship. ACF thanks the Royal Society. The authors thank the anonymous referee for helpful comments.

\label{lastpage}

\end{document}